\documentclass[twocolumn,prl,aps,superscriptaddress]{revtex4-1}
\usepackage[latin9]{inputenc}
\usepackage{mathtools}
\usepackage{amsbsy}
\usepackage{amssymb}
\usepackage{amstext}
\usepackage[unicode=true,pdfusetitle,
bookmarks=false,
breaklinks=false,pdfborder={0 0 1},backref=false,colorlinks=false]
{hyperref}
\hypersetup{
	bookmarksnumbered=false,bookmarksopen=false}
\makeatletter

\@ifundefined{textcolor}{}
{%
	\definecolor{BLACK}{gray}{0}
	\definecolor{WHITE}{gray}{1}
	\definecolor{RED}{rgb}{1,0,0}
	\definecolor{GREEN}{rgb}{0,1,0}
	\definecolor{BLUE}{rgb}{0,0,1}
	\definecolor{CYAN}{cmyk}{1,0,0,0}
	\definecolor{MAGENTA}{cmyk}{0,1,0,0}
	\definecolor{YELLOW}{cmyk}{0,0,1,0}
}

\newcommand{\beq}{\begin{equation}}
\newcommand{\eeq}{\end{equation}}
\newcommand{\beqa}{\begin{eqnarray}}
\newcommand{\eeqa}{\end{eqnarray}}

\usepackage{amsmath}
\usepackage{graphicx}
\usepackage{amssymb}
\usepackage{txfonts,color}
\makeatother
\begin{document}
	
\title{Breaking Adiabatic Quantum Control with Deep Learning}

\author{Yongcheng Ding}
\email{jonzen.ding@gmail.com}
\affiliation{International Center of Quantum Artificial Intelligence for Science and Technology (QuArtist) \\ and Department of Physics, Shanghai University, 200444 Shanghai, China}
\affiliation{Department of Physical Chemistry, University of the Basque Country UPV/EHU, Apartado 644, 48080 Bilbao, Spain}

\author{Yue Ban}
\affiliation{Department of Physical Chemistry, University of the Basque Country UPV/EHU, Apartado 644, 48080 Bilbao, Spain}
\affiliation{College of Materials Science and Engineering, Shanghai University, 200444
	Shanghai, China}

\author{Jos\'e  D. Mart\'in-Guerrero}
\email{jose.d.martin@uv.es}
\affiliation{IDAL, Electronic Engineering Department, University of Valencia,
	Avgda. Universitat s/n, 46100 Burjassot, Valencia, Spain}

\author{Enrique Solano}
\affiliation{International Center of Quantum Artificial Intelligence for Science and Technology (QuArtist) \\ and Department of Physics, Shanghai University, 200444 Shanghai, 
	China}
\affiliation{Department of Physical Chemistry, University of the Basque Country UPV/EHU, Apartado 644, 48080 Bilbao, Spain}
\affiliation{IKERBASQUE, Basque Foundation for Science, Plaza Euskadi 5, 48009 Bilbao, Spain}
\affiliation{IQM, Nymphenburgerstr. 86, 80636 Munich, Germany}

\author{Jorge Casanova}
\email{jcasanovamar@gmail.com} 
\affiliation{Department of Physical Chemistry, University of the Basque Country UPV/EHU, Apartado 644, 48080 Bilbao, Spain}
\affiliation{IKERBASQUE, Basque Foundation for Science, Plaza Euskadi 5, 48009 Bilbao, Spain}

\author{Xi Chen}
\email{xchen@shu.edu.cn}
\affiliation{International Center of Quantum Artificial Intelligence for Science and Technology (QuArtist) \\ and Department of Physics, Shanghai University, 200444 Shanghai, China}

\affiliation{Department of Physical Chemistry, University of the Basque Country UPV/EHU, Apartado 644, 48080 Bilbao, Spain}
\date{\today}

\begin{abstract}	

In the era of digital quantum computing, optimal digitized  pulses are requisite for efficient quantum control. This goal is translated into dynamic programming, in which a deep reinforcement learning (DRL) agent is gifted. As a reference, shortcuts to adiabaticity (STA) provide analytical approaches to adiabatic speed up by pulse control. Here, we select single-component control of qubits, resembling the ubiquitous two-level Landau-Zener problem for gate operation. We aim at obtaining fast and robust digital pulses by combining STA and DRL algorithm. In particular, we find that DRL leads to robust digital quantum control with operation time bounded by quantum speed limits dictated by STA. In addition, we demonstrate that robustness against systematic errors can be achieved by DRL without any input from STA. Our results introduce a general framework of digital quantum control, leading to a promising enhancement in quantum information processing.

\end{abstract}

\maketitle

\emph{Introduction.--} 
For many decades, quantum control is concerned with efficient manipulation of physical and chemical processes on the atomic and molecular scale, with various applications ranging from photochemistry to quantum information sciences~\cite{book-1,review}. Specifically, how to implement fast and robust qubit gates with externally controllable parameters is required for realizing universal fault-tolerant quantum computing in the physical platforms based on superconducting qubits and trapped ions~\cite{book-2}. Quantum error correction, for instance, has been developed to reduce the noise or imperfection coming from environment and control parameters themselves, in the implementation of applications for noisy intermediate-scale quantum (NISQ) gate-based computers~\cite{martinis15,TakitaPRL,Schoelkopf}. 

Two-level systems, called hereafter qubit systems, are the basic units of digital quantum computing. Thus, several studies have been devoted to produce distinct methods for precise quantum control of qubits with external fields. These include resonant pulses \cite{AE}, adiabatic passages~\cite{Revadiabatic}, composite pulses~\cite{Levitti,Chuang,vitanovprl,rongNC}, pulse-shape engineering~\cite{analytical,Steffen,PRL2013D}, and further optimizations~\cite{review,CanevaPRL,Gerhard,shaped,sugnypra13,sugnypra17,Rabitz}. Among these frameworks,  ``Shortcuts to adiabaticity" (STA)~\cite{RMP,STAreview} shares the concept  that break the adiabatic regime, and it leads to fast non-adiabatic state evolution by combining both merits of resonant pulses and adiabatic passages.
Specifically, inverse engineering, as one of the STA techniques, is applied to design the superadiabatic state evolution along the dynamical modes, emanated from the Lewis-Riesenfeld invariant, with appropriate boundary conditions \cite{chenprl104}. 
Thus, the freedom left in the inverse engineering 
further allows suppressing the inevitable systematic errors such as amplitude noise and dephasing noise, by incorporating other techniques of optimal control \cite{PRL2013D,inverse13,njp2012}, dynamical decoupling~\cite{Carlos}, and supervised machine learning~\cite{Sanders,manhong}. 


At the same time, along with the development of deep learning in many areas~\cite{MniharNature,Mniharxiv,alphago,silverarXiv}, deep reinforcement learning (DRL) has been used for different applications in physics~\cite{Carleo2017,Nagy,Hartmann,Vicentini,Iten}. DRL works as a promising method for searching optimal control pulses for fast and robust quantum state preparation~\cite{PNAS,Xinwang}, gate operation~\cite{Zhou}, and quantum Szilard engine \cite{Bergli}. More specifically, DRL enhances reinforcement learning, which is a key branch of classical machine learning widely applied to control tasks; this enhancement comes from the use of deep learning in key aspects of reinforcement learning. Recent works have studied the application of DRL to quantum control~\cite{Bukov,Niu,prawang1,prawang2,comm}. Therefore, we find that it is meaningful to compare DRL with STA for a better understanding of both. We believe that this study will lead to more feasible applications in manipulating superconducting transmon qubits~\cite{sigmazgate}, Bose-Einstein condensates~\cite{OliverNP}, and quantum dots \cite{WangXNC}, in which systematic errors, stochastic noise, and experimental constraint are of significance.

To this aim,  we explore fast and robust quantum control for qubit operation by combining STA and DRL methods. Strictly,
we focus on single-component control of qubits, being similar to 
the two-level Landau-Zener (LZ) problem~\cite{sigmazgate,OliverNP,WangXNC}, with the designed time-dependent frequency sweep. We show the first salient result: The smooth pulses of STA are analytically engineered with  clarified quantum speed limit (QSL), and further optimised with respect to various types of noises, imperfections, and physical constraints, particularly in the feasible experiments when lacking flexibility. More importantly, we look for such quantum control with DRL, benchmarking it by connecting STA. We find that DRL agent explores digital shortcuts for the same task, resulting in similar characteristics of robustness, when 
the operation time of STA is used as a hint. As an extension, we train the agent, which is the part of DRL in charge of taking control actions, without any input from STA for the efficient control by suppressing various systematic errors. In our numerical simulations, we further observe that DRL agent is capable of achieving efficient quantum control with satisfying features. We reckon that one can improve the performance of the framework by fine-tuning in an interactive DRL environment with quantum noise, resulting in the potential applications in Noisy Intermediate Scale Quantum (NISQ) systems.

\emph{Inverse engineering and optimization of STA.--} 
Consider the coherent manipulation of a single qubit, whose the Hamiltonian reads
\begin{equation} 
\label{HLZ}
H (t)=\frac{\hbar}{2}[\Omega\sigma_x+\Delta(t)\sigma_z],
\end{equation}
where the Rabi frequency $\Omega$ is fixed, while the detuning $\Delta(t)$ is time-varying. Equation~\eqref{HLZ} appears in, e.g., the Xmon transmon qubit \cite{sigmazgate}, in Bose-Einstein condensates within accelerated optical lattices~\cite{OliverNP}, and in quantum dot charge qubits~\cite{WangXNC}. According to the Lewis-Riesenfeld (LR) theory~\cite{LR}, one can construct a dynamical invariant $I(t) =   \frac{\hbar}{2}\Omega_0 \sum_{\pm} |\phi_\pm (t) \rangle \langle \phi_\pm (t)| $, where its eigenstates are $|\phi_+(t)\rangle=(\cos(\frac{\theta}{2})e^{-i\frac{\beta}{2}},\sin(\frac{\theta}{2}) e^{i\frac{\beta}{2}})^{\text{T}}$, and $|\phi_-(t)\rangle=(\sin(\frac{\theta}{2})e^{-i\frac{\beta}{2}},-\cos(\frac{\theta}{2})e^{i\frac{\beta}{2}})^{\text{T}}$, with $\Omega_0$ being an arbitrary constant frequency that keeps $I(t)$ in units of energy. Here the time-dependent angles $\theta\equiv\theta(t)$ and $\beta\equiv\beta(t)$ parametrise the trajectory of an evolving state on the Bloch sphere. The solution of time-dependent Schr\"{o}dinger equation is described by the superposition of $|\phi_\pm (t) \rangle$. More specifically, $|\Psi(t)\rangle=\sum_\pm c_\pm \exp(i\gamma_\pm)|\phi_\pm(t)\rangle$, with $c_n$ being constants, and the LR phases $\gamma_\pm$ are calculated as 
\begin{equation}
\label{gamma}
\gamma_\pm(t)=\pm\frac{1}{2}\int_0^t \left(\frac{\dot{\theta}\cot\beta}{\sin\theta}\right)dt'.
\end{equation} 
The condition, $dI(t)/dt\equiv\partial I(t)/\partial t+(1/i\hbar)[I(t),H(t)]=0$, yields the following auxiliary equations:
\begin{eqnarray}
\label{eq:dottheta}\dot{\theta}&=&-\Omega\sin\beta,
\\
\label{eq:dotbeta}\dot{\beta}&=&-\Omega\cot\theta\cos\beta+\Delta(t),
\end{eqnarray}
which enables the desired state to evolve along the dynamical mode, $|\phi_\pm (t)\rangle$. In previous works, a STA control framework facilitates optimization over errors and noise under symmetrical constraints \cite{njp2012,inverse13} with two tunable parameters -- i.e. Rabi frequency $\Omega$ and detuning $\Delta$ -- that hold Eqs.~\eqref{eq:dottheta} and~\eqref{eq:dotbeta}. However, in certain quantum platforms tunability on $\Omega$ and $\Delta$ is not available. For instance, in superconducting Xmon transmon qubits \cite{sigmazgate}  control on the detuning is preferred.

To adapt these requirements, we shall apply the inverse engineering method to design the angle parameter $\theta$ for tailoring the time-dependent detuning $\Delta(t)$. Accordingly, we substitute Eq.~\eqref{eq:dotbeta} and its derivative into Eq.~\eqref{eq:dotbeta}, leading to the following expression:
\begin{equation}
\label{eq:Delta}
\Delta(t)=-\frac{\ddot{\theta}}{\Omega\sqrt{1-\left(\frac{\dot{\theta}}{\Omega}\right)^2}}+\Omega\cot\theta\sqrt{1-\left(\frac{\dot{\theta}}{\Omega}\right)^2}.
\end{equation}
with constant Rabi frequency $\Omega$.  This allows us to drive the state evolution along one of dynamical mode,  $|\phi_+ (t) \rangle$, by single component within finite short time $T$, constrained by QSL. First of all, the following boundary conditions are imposed 
\beq
\label{bc-1}
\theta(0)=0, ~~	~~~ \theta(T)= \pi.
\eeq
These determine a qubit flip from $|0\rangle$ to $|1\rangle$. 
Secondly, the protocol can be optimised for cancelling systematic errors, as environmental fluctuations and deviations on the control parameters unavoidable in experimental scenario. To this end,  we consider the errors in Rabi frequency and detuning, i.e. $\Omega \rightarrow \Omega(1+\delta_{\Omega}) $ and 
$\Delta (t) \rightarrow \Delta(t) + \delta_{\Delta} $, and write down the transition probability, keeping the first-order term in the time-dependent perturbation theory \cite{supplementary}, 
\begin{equation}
\label{transition}
P=\frac{\hbar^2}{4}\left|\int_0^T\langle\Psi_-(t)| (\delta_{\Omega} \Omega \sigma_x + \delta_{\Delta} \sigma_z) |\Psi_+(t)\rangle\right|^2,
\end{equation}
with $|\Psi_\pm(t)\rangle = e^{ i  \gamma_\pm(t)}|\phi_\pm(t)\rangle $ being the two orthogonal dynamical modes of  the invariant. Plugging Eqs. (\ref{eq:dottheta}) and (\ref{eq:dotbeta}) into Eq.~(\ref{transition}), 
we obtain the following condition for error cancellation:
\begin{equation}
\label{errorcancelation}
\left|\int_{0}^{T}dt e^{i \eta(t)}\left(\delta_{\Delta}\sin\theta - i2\delta_{\Omega}\dot{\theta}\sin^2\theta\right)\right| = 0,
\end{equation}
where $\eta(t)=2\gamma_+(t)$, yielding $\dot{\eta}=\dot{\theta}\cot\beta/\sin\theta$ by combining with Eq. (\ref{gamma}). 

Inspired by Ref. \cite{PRL2013D}, a global phase $\eta(t)=2\gamma_+(t)$ in the integral~(\ref{errorcancelation}) is expanded as
\begin{equation}
\label{eq:eta}
\eta(t) = 2 \theta + \alpha_1 \sin (2 \theta) + \alpha_2 \sin (4 \theta) + ... + \alpha_n \sin (2 n \theta),
\end{equation} such that we get $\sin\beta=-1/\sqrt{1+4M^2\sin^2\theta}$, with $M=1+\sum_n n \alpha_n\cos(2n\theta)$. As a result, by solving Eq.~\eqref{eq:dottheta} with given coefficients $\alpha_n$ and the initial condition $\theta(0)=0$, one obtains the corresponding $\theta$ that evolves to $\theta(T)=\pi$, with $T$ bounded by QSL time as (see~\cite{Gerhard,Rabitz}),
\begin{equation}
\label{eq:qsl}
\Omega T=\int_0^\pi d\theta \sqrt{1+4 M^2\sin^2\theta}\geq\pi.
\end{equation}
This protocol allows robust qubit flipping from $|0\rangle$ to $|1\rangle$ with arbitrary series coefficients for pulse engineering. 
In principle, by introducing the free parameters $\alpha_n$ in Eq. (\ref{eq:eta}), one can nullify the above integral (\ref{errorcancelation}), such that the errors in both $\sigma_x$ and $\sigma_z$ terms can be simultaneously suppressed. For simplicity, here we may set $\delta_{\Omega}=0$ or $\delta_{\Delta} =0$, to independently discuss each error source, without presuming the ratio of error amplitudes. 
We find out that $\Omega$-error and $\Delta$-error can be eliminated with only first-order expansion of $\eta(t)$, resulting in $\alpha_{1}=-1$ and $\alpha_{1}=-1.74$, respectively.

\begin{figure}
	\includegraphics[width=8.5cm]{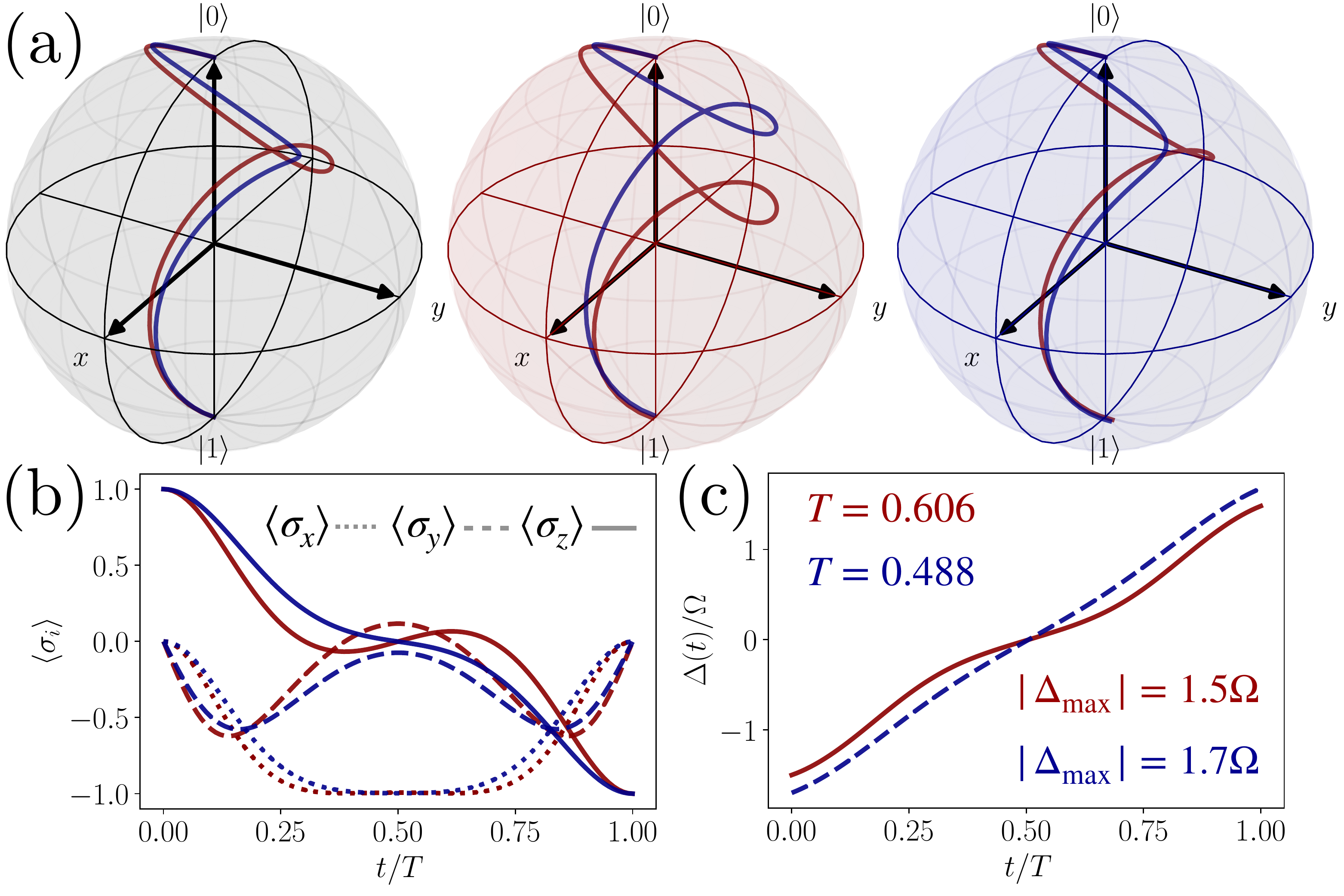}
	\caption{\label{fig:a}(a) STA of qubits on Bloch spheres, where the continuous pulses of LZ type are designed from Eq.~\eqref{eq:Delta} by using the \textit{Ans\"atz} of $\theta(t)$~\eqref{eq:ansatz}. With $a=0.604$ (red curve) and $0.728$ (blue curve), the qubit without any errors (gray sphere) is flipped from $|0\rangle$ to $|1\rangle$, being robust against $\Delta$- and $\Omega$-errors, respectively, which is proved by numerical simulations under the amplitudes of $\lambda_\Delta=\pm0.1\Delta_{\text{max}}$ (red and blue curves on red sphere) and $\lambda_\Omega=\pm0.1\Omega$ (red and blue curves on blue sphere). (b) Evolution of expectations on different directions, corresponding to the trajectory on gray sphere, by minimizing $\Delta$-error (red curve) or $\Omega$-error (blue curve). We fix Rabi frequency $\Omega=20\times2\pi$ MHz, resulting in operation time $T=60.6$ ns and  $T=48.8$ ns for STA protocols canceling previous errors. (c) The corresponding smooth pulses are depicted as well.}
\end{figure}

While robust quantum control can be achieved within the preceding framework, we notice that one can hardly predict the shape of the detuning as well as its adjustable range. For example, the detuning against $\Delta$-error has the maximum amplitude of more than $3\Omega$ with abrupt changes at the edges of the operation (see Supplementary Material ~\cite{supplementary}). For more feasible implementations, we prefer smooth controls with a detuning $\Delta(t)$ that does not oscillate drastically. Thus, we propose the following \textit{Ans\"atz} for $\theta$
\begin{eqnarray}
\label{eq:ansatz}
\theta(t)=\frac{\Omega T}{a}\left[as-\frac{\pi^2}{2}(1-s)^2
+\frac{\pi^2}{3}(1-s)^3+\cos(\pi s)+A\right],~~~~~~~~~
\end{eqnarray}
where $s=t/T$, $A=\pi^2/6-1$, and $T={-\pi a}/[{(2-a-\pi^2/6)\Omega}]$ are found by using the boundary conditions~(\ref{bc-1}), while $a>2-\pi^2/6$ being a free parameter. This \textit{Ans\"atz} leads to a detuning $\Delta(t)$ that grows almost linearly during the operation time, resembling the original LZ scheme, but with finite values at $t=0$ and $t=T$ (see Fig.~\ref{fig:a} and Supplementary Material~\cite{supplementary}). Accordingly,  this protocol provides with $a=0.604$ and $0.728$ for robust qubit flipping against $\Delta$- and $\Omega$-errors, respectively, resulting in the operation time $T=60.6$ ns and  $T=48.8$ ns, of the same order of characteristic gate time for superconducting Xmon transmon qubit \cite{sigmazgate}.  

\begin{figure}
	\includegraphics[width=8.56cm]{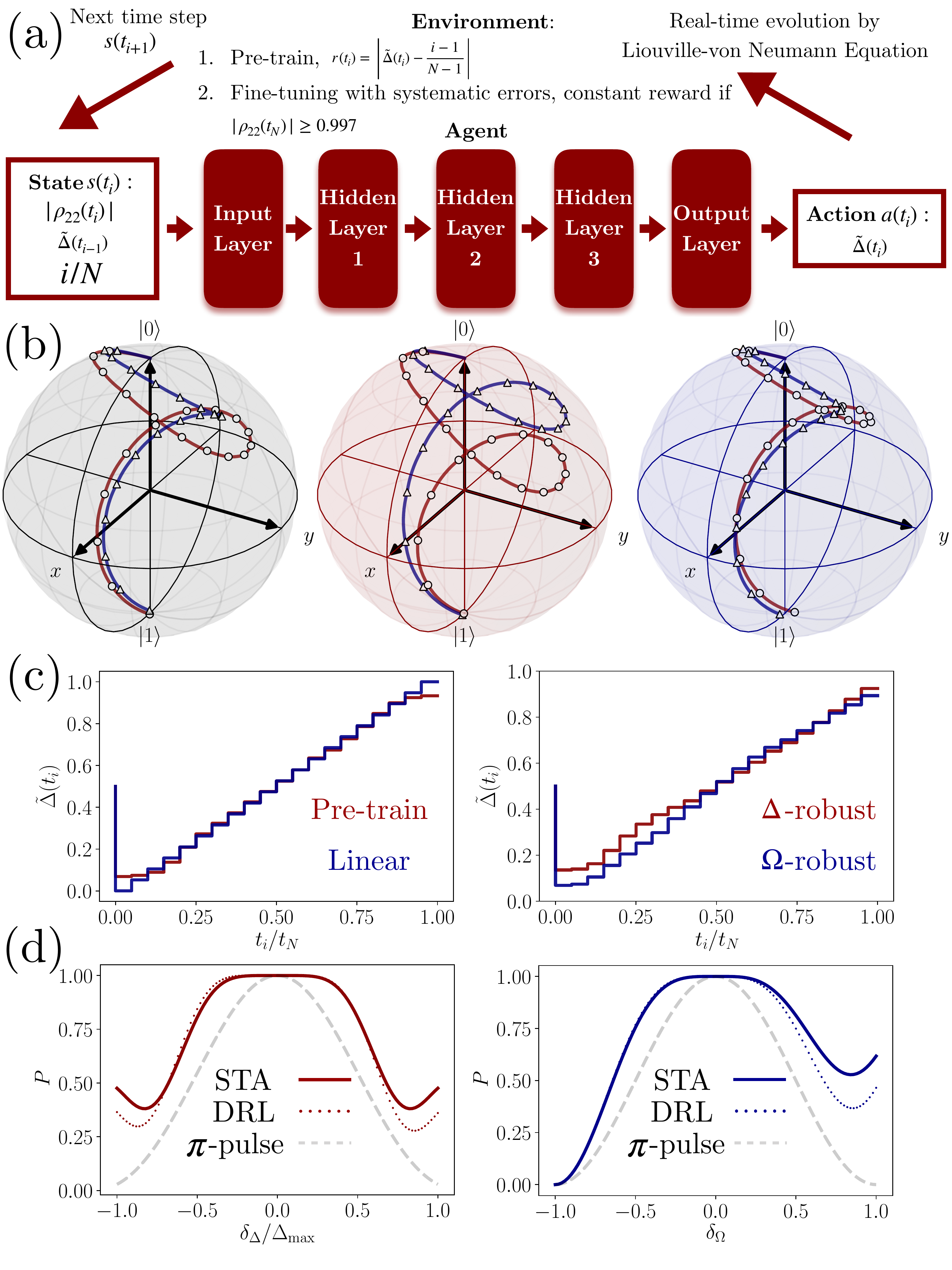}
	\caption{\label{fig:b}(a) Scheme of DRL approach to quantum control with LZ scheme for one time step in training. An ANN (Agent) of three hidden layers observes a state, which encodes physical information of the qubit. An action is outputted, evolving the system for one time step, resulting in its state of next time step. Environment rewards or punishes the agent by artificially designed reward function, enabling the agent to learn by accumulation of them. (b) State evolution of the qubit on Bloch spheres, driven by digital pulses from DRL within 20 time steps, where the parameters are the same as those in Fig.~\ref{fig:a}. (c) Renormalized detuning pulses after pre-training for control of LZ type, and fine-tuning according to systematic errors and populations. Following Fig.~\ref{fig:a}(c), $\Delta_{\max}$ are set to $1.5\Omega$ and $1.7\Omega$ for $\Delta/\Omega$-error, respectively. (d) Final population of state $|1\rangle$ versus relative systematic errors. Protocols designed by STA and obtained from DRL are both robust against systematic errors with similar feature, comparing to resonant flat $\pi$-pulse as time-optimal solution.}
\end{figure}

\emph{Deep reinforcement learning.--} 
Even though STA has enabled the fast and robust control, we consider other numerical methods for more complicated cases, e.g., if we are only allowed to drive the quantum states with a given number $N$ of detuned pulses within a fixed time. This task is indeed combinational optimization (i.e. maximizing robustness with optimal configuration of discretized pulses) which is equivalent to dynamic programming, namely, a decision problem of multiple steps. Although the complexity of dynamic programming grows exponentially with the number of steps, one can still approximately solve it, e.g., with an artificial neural network (ANN) approach, which naturally leads us to the concept of DRL. Actually, when one talks about DRL two main approaches arise. The first one is based on the use of deep learning to approximate the dynamic programming solution. The second approach deals with the so-called deep policy networks, i.e., the ability to test many different control systems in parallel. We focus on the former approach. In this framework, the assumption of DRL is that, there exists an optimal policy $\pi$, giving an action $\textbf{a}(t_i)$ for any observable state $\textbf{s}(t_i)$ to complete a certain task in a system. This state-action relation can be characterized by a function $\pi(\textbf{a}|\bf{s})$, which can be approximated by a deep ANN. State $\textbf{s}(t_i)$ is encoded in input variables within zero to one, being observed by the agent ANN. After propagations between layers and nonlinear activations of the ANN nodes, output layer gives an action $\textbf{a}(t_i)$, which evolves the system to next state $\textbf{s}(t_{i+1})$ within one timestep.



An environment consists of these equations, governing the evolution, as well as rewarding the agent. Optimizers tune parameters of the ANN according to rewards, leading to a well-trained agent to provide optimized actions for completing the task. Thus, we notice that the concept of inverse engineering from STA, i.e., choosing an \textit{Ans\"atz} to procure the protocol via auxiliary equations, is similar to DRL. One should design a reward function $r(\textbf{s},t_i)$ to educate the agent. An adequate reward function accelerates the convergence of DRL algorithm, preventing the agent from getting stuck into trivial solutions or cheating by repetitive actions.

\begin{figure}
\includegraphics[width=8.5cm]{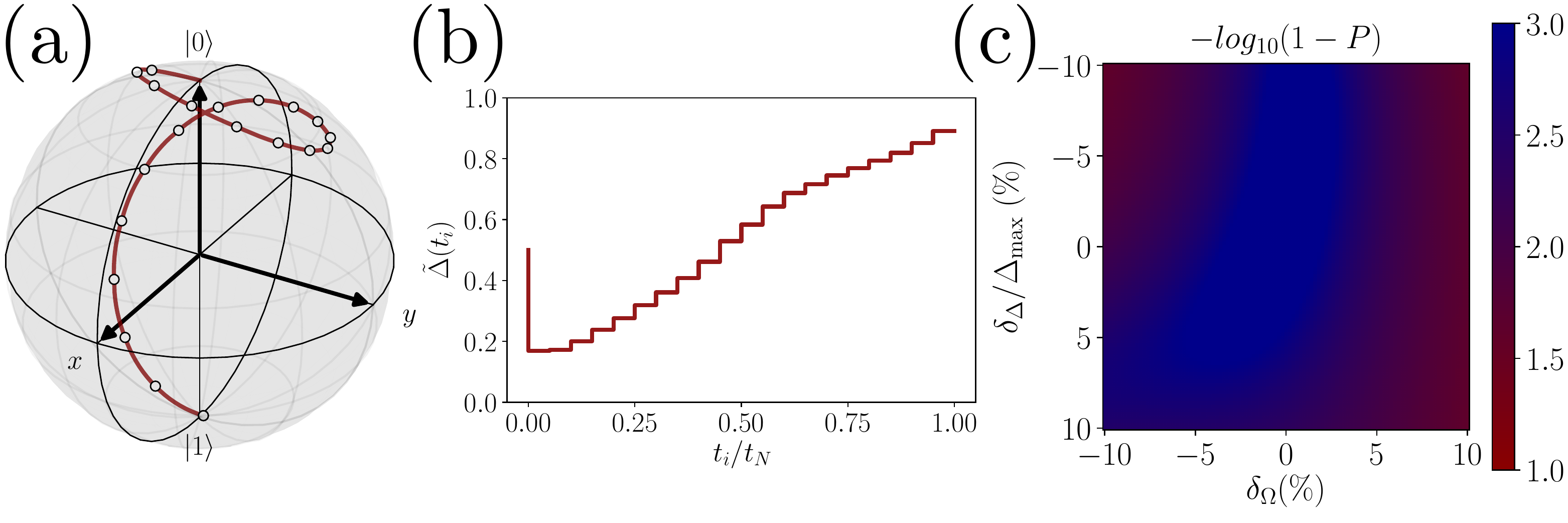}
\caption{\label{fig:c}(a) State evolution during an arbitrary time $t_N=55$ ns in an ideal system. (b) Digital pulses given by DRL, $\Delta_\text{max}$ is chosen as $1.6\Omega$ without knowledge of STA. (c) Deviation of population under both $\Omega$- and $\Delta$-errors.}
\end{figure}

In our practice of DRL, we renormalize the tunable detuning range $[-\Delta_{\text{max}},\Delta_{\text{max}}]$ into $\tilde{\Delta}\in[0,1]$, which is the encoded action at time step $t_i$: $\tilde{\Delta}(t_i)=(\Delta(t_i)+\Delta_{\text{max}})/2\Delta_{\text{max}}$. The state consists of the element of the density matrix $|\rho_{22}(t_i)|$, renormalized detuning  $\tilde{\Delta}(t_{i-1})$ as the action of last time step, and the current system time $i/N$. Here we use the Liouville-von Neumann equation instead of Schr\"{o}dinger equation for further generalization to the Lindblad master equation with quantum noise involved elsewhere. 
The training strategy of the agent is shown in Fig.~\ref{fig:b}(a), where extensive details regarding this strategy can be found in the Supplementary Material~\cite{supplementary}. We find that DRL agents converge to (sub-)optimal solutions by approximating the policy that maximizes the artificial reward. This enables DRL to explore various types of quantum control, requiring adequate reward function for our task. For example, we notice that Proximal Policy Optimization (PPO) rapidly learns the time-optimal solution, i.e. resonant flat $\pi$ pulse, with a trivial reward function $r(t_i)=|\rho_{22}(t_i)|-1$. However, for LZ-type control, we pre-train the agent with $r(t_i)=-|\tilde{\Delta}(t_i)-\frac{i-1}{N-1}|$, rewarding linear growth of detuning. Pre-training can filter other strategies and accelerate the convergence as well. Later, we reward the agent by a constant for fine-tuning, if $|\rho_{22}|> 0.997$ at the final time step with random systematic errors.

We firstly investigate if the DRL agent learns digital quantum control resembling STA, [see Fig.~\ref{fig:a}(c)], with operation time $T=60.6$ ns and $48.8$ ns split equally by using 20 pulses as the only hint. The control calculated by STA eliminates the error transition, which bounds the upper limit of robustness. Thus, the performance of DRL can be easily benchmarked. We find out that the DRL agent manages to flip the qubit against systematic errors by digital pulses [see Fig.~\ref{fig:b}(b) and (c)], which are not the coarse-grained analog controls.

In Fig.~\ref{fig:b}(d), we compare the robustness of STA, DRL and flat $\pi$-pulse against $\Delta/\Omega$-errors. The agent discovers digital quantum control with the same feature of STA, which is quite satisfying for approaching the theoretical maximum of robustness. Inspired by this preliminary result, we further employ DRL for the sake of searching robust digital control against both $\Omega$- and $\Delta$-errors. In this scenario, DRL is more straightforward since the inverse engineering from STA does not work perfectly even with more free parameters, depending on the certain proportion of $\delta_\Delta$ and $\delta_\Omega$. With the same training strategy, we educate the DRL agent for qubit flipping against both types of errors, which is shown in Fig.~\ref{fig:c}. It is worthwhile to mention that, we set operation time and tunable range of detuning without any knowledge from STA. The agent learns its goal, resulting in populations exceeding $0.99$ within $\delta_\Delta/\Delta_{\max},\delta_\Omega\in[-0.1,0.1]$.

\emph{Discussion.--} 
We deem the research presented here as novel and impacting. Although one may argue that other numerical algorithms like GRAPE and CRAB are also capable of completing similar tasks, in our practice, we perceive that these gradient algorithms have their  limitations. Some of them easily get stuck to local minima, being far from optimal solutions because of initial configurations. By contrast, it is proved that the global minimum can be achieved in ANN with gradient descent~\cite{global1,global2}. Studies also verify that any complex ANN can be reduced to one with much smaller sizes without loss of performance~\cite{lottery}, massively saving training time. These theoretical researches will continuously improve DRL, widening its application in quantum control. We also emphasize the critical issue that we are far from exploiting the power of DRL in this work because of physical constraints. In other scenarios of applied DRLs, states for agents are easily to be observed. For example, states of Go, RTS games, and automatic driving are already digitized during information collection processes. However, precise observation of states in quantum control destructs the system immediately, requiring enormous copies for learning by real devices or controlling an unknown system by trained agents. Therefore, we use digital pulses given by the agent in an ideal environment for all systematic errors. In other words, we forbid the agent to observe any state during evaluation, which is also for a fair comparison between fixed-STA and DRL. In spite of the constraint we imposed, the cost of observing the state for DRL agent in quantum control can be reduced by experimental strategies inspired by machine learning~\cite{mario116,marioreview,rqi}. If an agent is allowed to observe the state before each time step, it should dynamically change the action according to optimized policies as it does for other tasks. This improvement also requires a precise measurement of the quantum state. Otherwise, the performance can even be worse than fixed quantum control.

\emph{Conclusion.--} 
By comparison, we figure out that STA provides a well-optimized analytical method for designing fast and robust quantum control, in terms of inverse engineering with parameter variations. Beyond that, it enhances the performance of DRL, providing knowledge of optimal evolving time, accelerating its convergence as well. On the other hand, DRL also shows its capability of learning physics with artificially designed reward functions. The agent obtains digital pulses with complex constraints, which prominently eliminate systematic errors. Pulses designed for more complicated cases are also satisfying even without field knowledge. Last but not least, our DRL framework can be extended to multiqubit systems \cite{YChenprl2014,Gellerpra} and quantum noise without many efforts, for the application of robust control optimization in quantum algorithms \cite{PRX2017,Dongarxiv} based on NISQ devices. Moreover, agent-training in interactive environment with real quantum devices as fine-tuning can lead to a significant enhancement, if experimental time cost is acceptable.

\begin{acknowledgements}
This work is partially supported from NSFC (11474193), STCSM (2019SHZDZX01-ZX04, 18010500400 and 18ZR1415500), Program for Eastern Scholar, QMiCS (820505) and OpenSuperQ (820363) of the EU Flagship on Quantum Technologies,  EU FET Open Grant Quromorphic, Spanish Government PGC2018-095113-B-I00 (MCIU/AEI/FEDER, UE), and Basque Government IT986-16.
X. C. acknowledges Ram\'on y Cajal program (RYC-2017-22482).
J. C. also acknowledges the Ram\'{o}n y Cajal program (RYC2018-025197-I) and support from the UPV/EHU through the grant EHUrOPE. 

\end{acknowledgements}

\pagebreak
\widetext
\begin{center}
	\textbf{ \large Supplemental Material: \\ Breaking Adiabatic Quantum Control with Deep Learning}
\end{center}

\setcounter{equation}{0} \setcounter{figure}{0} \setcounter{table}{0}
\global\long\def\thefigure{S\arabic{figure}}
\global\long\def\bibnumfmt#1{[S#1]}
\global\long\def\citenumfont#1{S#1}

\section{Optimization of shortcuts to adiabaticity}
We have a qubit described by a two-level system
\begin{equation}
H(t)=\frac{\hbar}{2} [\Omega \sigma_x +\Delta(t) \sigma_z],
\end{equation}
with Landau-Zener (LZ) scheme, where only-$\sigma_z$-control is allowed due to experimental constraints. Parameterized Lewis-Riesenfeld (LR) invariant gives \cite{SMinverse13}
\begin{equation}
I(t)=\frac{\hbar}{2}\Omega_0 (  \sin\theta \cos\beta \sigma_x +\sin\theta \sin\beta \sigma_y +   \cos\theta \sigma_z),
\end{equation}
where its eigenstates read
\begin{equation}
|\phi_+(t)\rangle=\begin{pmatrix}
\cos(\frac{\theta}{2})e^{-i\frac{\beta}{2}}\\
\sin(\frac{\theta}{2})e^{i\frac{\beta}{2}}
\end{pmatrix}, ~~~~ |\phi_-(t)\rangle=\begin{pmatrix}
\sin(\frac{\theta}{2})e^{-i\frac{\beta}{2}}\\
-\cos(\frac{\theta}{2})e^{i\frac{\beta}{2}}
\end{pmatrix}.
\end{equation}
The solution of time-dependent Schr\"{o}dinger equation can be described by the superposition of dynamical modes
\begin{equation}
|\Psi(t)\rangle=\sum_\pm c_\pm\exp(i\gamma_\pm)|\phi_\pm(t)\rangle,
\end{equation}
where the LR phases are 
\begin{eqnarray}
\label{supeq:lr}
\gamma_\pm(t)\frac{1}{\hbar}\int_0^t\langle\phi_\pm(t')|i\hbar\frac{\partial}{\partial t'}-H(t')|\psi_\pm(t')\rangle dt' =\pm\frac{1}{2}\int_0^t \left(\frac{\dot{\theta}\cot\beta}{\sin\theta}\right)dt'.
\end{eqnarray}
In this case, the time evolution operator is represented as
\begin{equation}
\hat{U}(T,t)=|\Psi_{-}(T)\rangle\langle\Psi_+(t)|+|\Psi_-(T)\rangle\langle\Psi_-(t)|,
\end{equation}
with $|\Psi_{\pm}(t) \rangle =\exp(i\gamma_\pm)|\phi_\pm(t)\rangle$. 
Next, we consider a general systematic errors $H'$, and write down the perturbated state evolution at $t=T$, keeping second-order perturbation theory up to $O(\lambda^2)$,
\begin{eqnarray}
|\tilde{\Psi}(T)\rangle=|\Psi(T)\rangle-\frac{i}{\hbar}\int_0^T\hat{U}(T,t)H'(t)|\Psi(t)\rangle\nonumber-\frac{1}{\hbar^2}\int_0^Tdt\int_0^tdt'\hat{U}(T,t)H'(t)\hat{U}(t,t')H'(t')|\Psi(t')\rangle+\cdots.
\end{eqnarray}
Thus, after neglecting the higher-order perturbative terms, the final probability to be in $|1\rangle$ is calculated as,
\begin{eqnarray}
P \approx 1-\frac{1}{\hbar^2}\left|\int_0^Tdt\langle\Psi_-(t)|H'|\Psi_+(t)\rangle\right|^2.
\end{eqnarray}
As a consequence, we can specify the transition probability induced by the systematic errors,
\begin{eqnarray}
\label{supeq:P}
P=\frac{1}{4}\left|\int_0^Tdt e^{i\eta(t)} (\delta_{\Delta} \sin\theta  - i 2\delta_{\Omega}\dot{\theta}\sin^2\theta)\right|^2,
\end{eqnarray}
with $\eta(t)=2\gamma_+$, when we consider the errors in Rabi frequency and detuning, i.e. $\Omega \rightarrow \Omega(1+\delta_{\Omega}) $ and 
$\Delta (t) \rightarrow \Delta(t) + \delta_{\Delta} $, respectively. The simultaneous error cancellation requires the presumption of ratio of error amplitudes, i.e. $\delta_{\Delta}/\delta_{\Omega}$. Therefore, one can simple  consider the individual systematic error in Rabi frequency and detuning, by imposing $\delta_{\Delta}=0$ or $\delta_{\Omega}=0$ in Eq. (\ref{supeq:P}).

\begin{figure}
	\includegraphics[width=15cm]{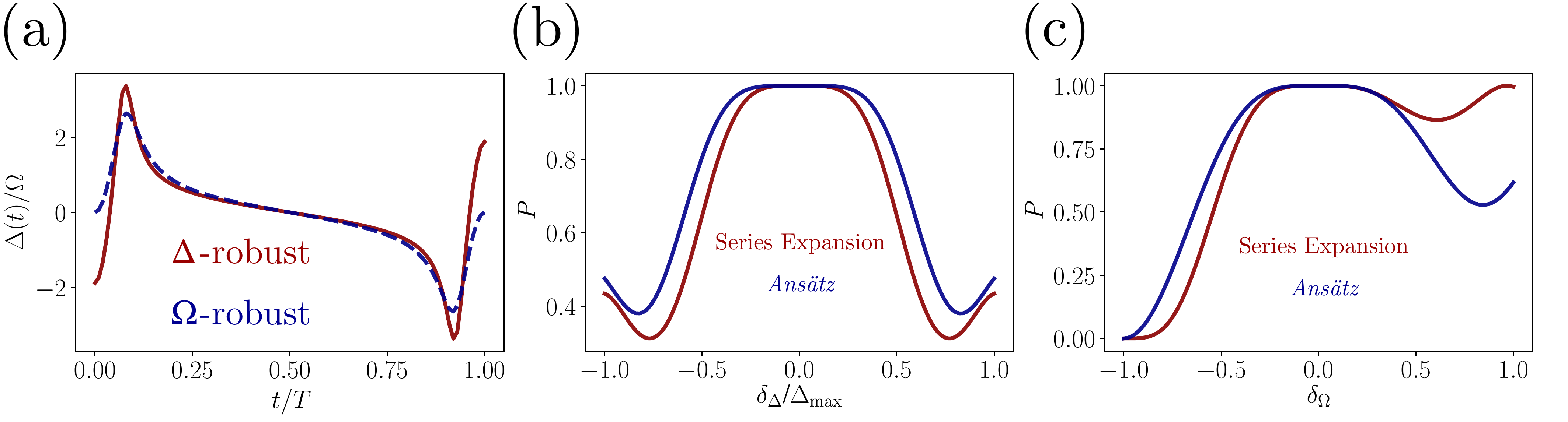}
	\caption{\label{fig:sup0}(a) Detuning pulses derived from Eq.~\eqref{supeq:Delta}, with maximal detuning amplitude of $3\Omega$ and abrupt changes at the edges. Thus, we introduce the \textit{Ans\"atz}~\eqref{supeq:ansatz} for more feasible detuning pulses in Fig.~\ref{fig:a}(c) The operation time of quantum control via series expansion gives $T=60.2$ ns and $53.9$ ns, being around those in the manuscript as $T=60.6$ ns and $48.8$ ns. (b) and (c) Population versus relative systematic errors, with the same definition and parameters of the manuscript.}
\end{figure}

In order to suppress the systematic errors by nullifying the integral (\ref{supeq:P}), we expand the global phase, $\eta(t)$ \cite{SMPRL2013D},  
\begin{equation}
\eta(t) = 2 \theta + \alpha_1 \sin (2 \theta) + \alpha_2 \sin (4 \theta) + ... + \alpha_n \sin (2 n \theta).
\end{equation}
By taking the first derivative on both sides of $\eta(t)$ and combining with Eq.~\eqref{supeq:lr}, we obtain
\begin{eqnarray}
\dot{\eta}&=&\dot{\theta}\cot\beta/\sin\theta,
\end{eqnarray}
where 
\begin{equation}
\label{supeq:sinbeta}
\sin\beta=-\frac{1}{\sqrt{1+4M^2\sin^2\theta}},
\end{equation}
with $M=1+\sum_n\alpha_n\cos(2 n\theta)$ ($n=1,2,3...$). Thus, the QSL time $T$ is determined by integrating Eq.~\eqref{eq:dottheta}:
\begin{eqnarray}
\label{supeq:T}
\Omega T=\int_0^\pi d\theta\sqrt{1+4M^2\sin^2\theta}\geq\pi,
\end{eqnarray}
with a lower bound of $\pi$ as given by resonant $\pi$-pulse. In this way, we solve Eq.~\eqref{eq:dottheta} with initial condition $\theta(0)=0$:
\begin{equation}
\dot\theta=\frac{\Omega}{\sqrt{1+4M^2\sin^2\theta}},
\end{equation}
ensuring $\theta(T)=\pi$ because of QSL time (\ref{supeq:T}). The evolution of $\theta$ and operation time $T$ is only determined by the series coefficients $\alpha_n$, which are numerically optimized to minimize error cancellation condition for suppressing $\Delta/\Omega$-error or both of them. By transforming Eq.~\eqref{supeq:sinbeta}, we get the following expression of $\beta$:
\begin{equation}
\beta=-\text{arccot}(2M\sin\theta),
\end{equation}
with its derivative being
\begin{equation}
\label{supeq:dbeta}
\dot{\beta}=\frac{-4\dot{\theta}\sin\theta\left[\sum_n n^2\alpha_n \sin\left(2 n\theta\right)\right]+2M\dot{\theta}\cos\theta}{1+4M^2\sin^2\theta}.
\end{equation}
We calculate the expression of detuning pulses by substituting Eq.~\eqref{supeq:dbeta} and~\eqref{supeq:sinbeta} into Eq.~\eqref{eq:dotbeta}:
\begin{eqnarray}
\label{supeq:Delta}
\Delta(t)=\frac{-4\dot{\theta}\sin\theta\left[\sum_n n^2\alpha_n\sin\left(2n\theta\right)\right]+2M\dot{\theta}\cos\theta}{1+4M^2\sin^2\theta}+\frac{2M\Omega}{\sqrt{1+4M^2\sin^2\theta}}.
\end{eqnarray}
As we mentioned in the main text, one can obtain the robust quantum control against the individual systematic error, see Fig.~\ref{fig:sup0}(a), e.g., $\alpha_1=-1.74$ for $\Delta$-error and $\alpha_1=-1$ for $\Omega$-error.

However, the shape of pulses given by Eq.~\eqref{supeq:Delta} changes abruptly, requiring a larger amplitude as well. Although pulse shape can be further optimized by introducing higher-order expansion, the search of optimal series coefficients that minimize the error cancellation condition is notorious since it grows in dimension. Meanwhile, it is also impossible to choose the optimal smooth pulse automatically from infinite sets of series coefficients which all cancel the transition.

For a more feasible experimental implementation, we prefer smooth control so that the detuning does not oscillate drastically.
The way out is to construct an \textit{Ans\"atz} of $\theta$ for satisfying boundary conditions of higher order:
\begin{equation}
\dot{\theta}(0)=\Omega,~\dot{\theta}(T)=\Omega,~\ddot{\theta}(0)=0,~\ddot{\theta}(T)=0,
\end{equation}
by combining polynomial and trigonometric functions as
\begin{eqnarray}
\label{supeq:ansatz}
\theta(t)=\frac{\Omega T}{a}\left[as-\frac{\pi^2}{2}(1-s)^2
+\frac{\pi^2}{3}(1-s)^3+\cos(\pi s)+A\right],
\end{eqnarray}
where $s=t/T$, $A=\pi^2/6-1$ and $T={-\pi a}/[{(2-a-\pi^2/6)\Omega}] \geq \pi/\Omega$ with $a>2-\pi^2/6$ being a free parameter. By searching $a$ that nullifies Eq. (\ref{supeq:P}), one obtains the robust smooth detuning pulses
with $a=0.604$ and $a=0.728$ in Fig~\ref{fig:a}, featuring  similar robustness as those given by series expansion [see Fig~\ref{fig:sup0}(b) and (c)].

\section{Optimal Control and Quantum Speed limit}

\begin{figure}
\includegraphics[width=14cm]{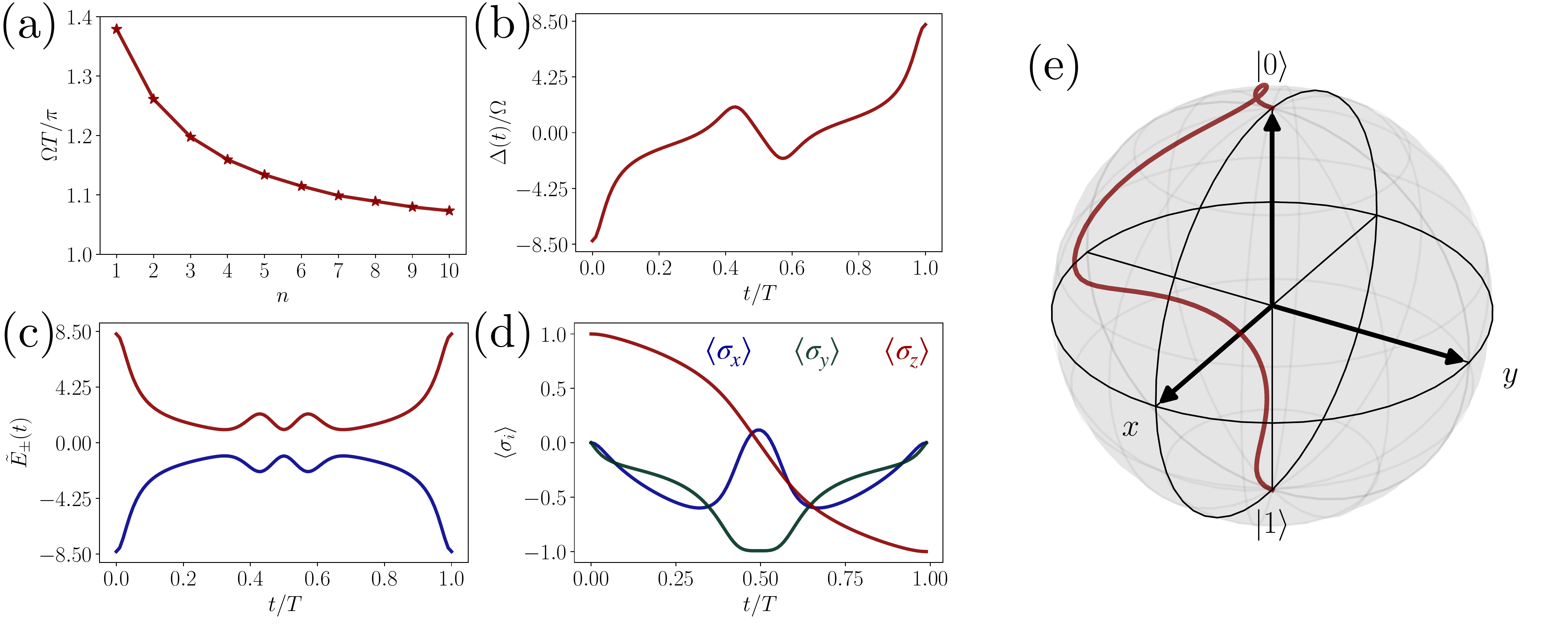}
\caption{\label{fig:sup1}(a) QSL time versus $n$. (b) Detuning with quasi-time-optimal scheme, when $\alpha_1=1.06$ minimizes QSL time, $T=34.5$ ns, by first-order expansion, for $n=1$. (c) Rescaled dimensionless instantaneous eigenvalues of energy defined by $\tilde{E}_\pm(t)=\pm\sqrt{\Omega^2+\Delta(t)^2}/\Omega$. (d) Expectations on different directions versus evolving time. (e) Evolution of quantum state on the Bloch sphere.}
\end{figure}

Now we focus on the time-optimal control and QSL time for the two-level qubit system with LZ scheme. In general, the fastest protocol is given by bang-off-bang pulse, in the unconstrained case when $\Delta_{\max}$ is sufficient, i.e.
\begin{equation}
\Delta(t) =\left\{
\begin{array}{cc}
-\Delta_{\max}, & t=0 \\
0, & 0<t<T \\
\Delta_{\max}, & t=T
\end{array}
\right.,
\end{equation}
such that the QSL time can be founded by \cite{SMGerhard}
\begin{equation}
\Omega T_{QSL} = \arccos(|f_0 i_0| +|f_1 i_1|) \geq \pi,
\end{equation}
when the initial and final state are $|\Psi(0) \rangle=i_0 |0\rangle+ i_1 |1\rangle $ and $|\Psi(T) \rangle=f_0 |0\rangle+ f_1 |1\rangle $, yielding $T_{QSL} = \pi/\Omega$ for the ideal qubit flipping. However, the sudden change of detuning makes the experimental implementation difficult or unfeasible. Here we investigate the minimal time bounded by QSL time, see Eq. (\ref{supeq:T}), with various order expansion of the global phase $\eta(t)$. We notice that the minimum of QSL converges to $\pi$ with the growth of $n$ as follows: (1, 4.33), (2, 3.96), (3, 3.76), (4, 3.64), (5, 3.56), (6, 3.5), (7, 3.45), (8, 3.42), (9, 3.39), (10, 3.37)..., where we show the time-optimal control in Fig.~\ref{fig:sup1}. Here, the required detuning range is much larger than those for nullifying error sensitivity, which clarifies the cost of STA optimization, and the trade-off between time and robustness.

\begin{figure}
\includegraphics[width=14cm]{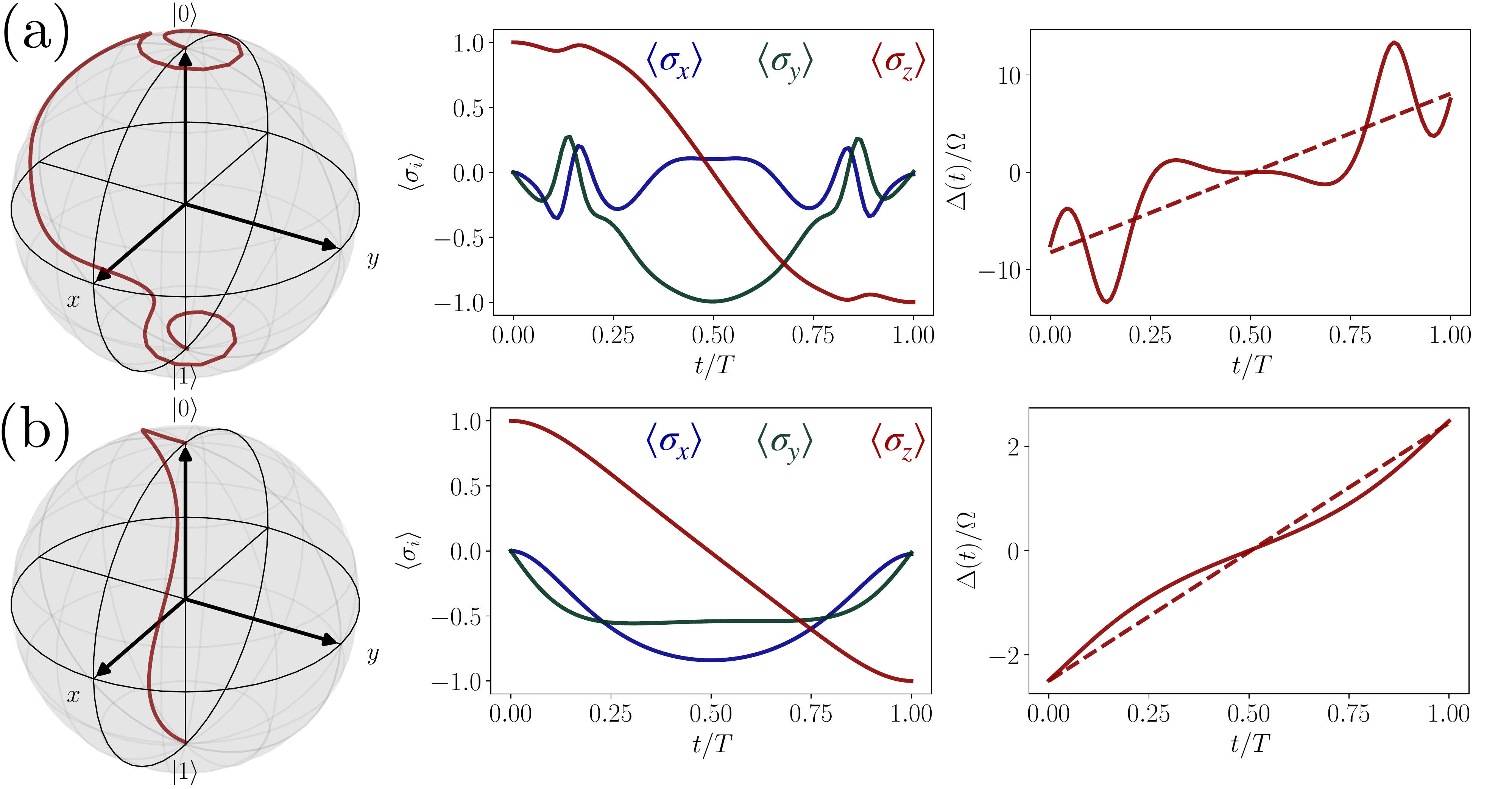}
\caption{\label{fig:sup2} Time-optimal solutions with LZ scheme found by GRAPE, where the initial $|\Delta|_\text{max}=$ (a) 8.24 (the maximal detuning when $n=1$ in Eq. (\ref{supeq:T})), and (b) 2.5, we notice that the result of GRAPE is unpredictable, being too sensitive to initialization.}
\end{figure}

Moreover, we also implement GRadient Ascent Pulse Engineering (GRAPE) algorithm which is considered as the state-of-the-art algorithm for designing digital pulses. GRAPE formulates the quantum control problem
\begin{equation}
H(t)=H_d+\sum_{j=1}^M u_j(t)H_j,
\end{equation}
where $H_d$ is the drift Hamiltonian, $H_j$ is the control Hamiltonian. The algorithm drives the quantum system to a target state with $M$ digital pulses, where amplitudes are denoted by $u_j(t)$. Unitary operator reads as
\begin{equation}
U(T,0)=\exp\left[-\frac{i}{\hbar}\int_0^T\frac{H(t)}{\hbar} dt
\right],
\end{equation}
which is discretized by
$
U(T,0)=\Pi_{k=1}^{M}U(t_k),
$
with $U(t_k)=\exp\left[-i\Delta t\frac{H(t_k)}{\hbar}\right]$,
evolving the system for a timestep of $\Delta t=T/M$. Amplitudes are expressed by an $M$-dimensional vector $|u_r\rangle$, updating by the gradient of figure-of-merit $f$ with learning rate of $\epsilon$
\begin{equation}
|u_{r+1}\rangle=|u_r\rangle+\epsilon\mathcal{H}^{-1}|\nabla f_r\rangle,
\end{equation}
with $\mathcal{H}$ to be the Hessian matrix. This formalism can be approximated since calculating  inversion of Hessian matrices is too time-consuming. In our practice, we design pulses with LZ-type scheme by initializing amplitudes of each pulse linearly. However, we find out that GRAPE is highly sensitive to initial configurations of pulses like other gradient algorithms (see Fig.~\ref{fig:sup2}). Moreover, GRAPE requires more quantum resources for discovering pulses in systems with unknown systematic errors. One cannot include these errors by codes, but giving feedbacks from the quantum device iteratively, practicing a quantum-classical training. By contrast, a DRL agent can be trained classically, saving quantum resources, giving pulses without observations of quantum states.

\section{Proximal Policy Optimization and its implementation}

In our practice, we construct an ANN made up of three hidden layers with 32 neurons each and ReLU as activation function. There are only three neurons in the input layer, which characterize the element of density matrix at time step $t_i$ by $|\rho_{22}(t_i)|$, renormalized detuning of last time step by $\tilde{\Delta}(t_{i-1})$, and the current system time by $i/N$. Output of the ANN is a renormalized detuning $\tilde{\Delta}(t_i)=(\Delta(t_i)+\Delta_{\text{max}})/2\Delta_{\text{max}}$ at time step $i$, by assuming that detuning is tunable between $[-\Delta_{\text{max}},\Delta_{\text{max}}]$. We notice that introducing current system time into ANN accelerates the learning process, since agents should be aware of how long has the system been evolving in a task for transporting within a fixed duration for deciding the action for the next time step. 

We implement Proximal Policy Optimization (PPO)~\cite{ppo} for our DRL approach since it performs comparably or better than state-of-the-art algorithms. For example, Trust Region Policy Optimization (TRPO)~\cite{trpo} performs well in continuous control problem, being robust against hyper-parameters. However, it requires a larger batch size for sampling, being more time consuming or even breaking down in high-dimensional problems. It also meets the difficulties when policy and reward function share parameters. Instead of TRPO, PPO finds a balance between sample complexity and ease of tuning, updating the policy by a relative small deviation from the previous one. We notice that PPO rapidly learns the time-optimal solution of single-component control with a trivial reward function of $r(t_i)=|\rho_{22}(t_i)|-1$ if we do not constrain the evolving time to a certain value, resulting in a constant zero solution, i.e., conventional flat $\pi$-pulse. For verifying if DRL can explore desired pulses against systematic errors, we start from retrieving digital pulses within evolving time calculated by STA in Fig.~\ref{fig:a}. We constrain the tunable range of detuning within the same region of those calculated by STA, as well as the evolving time. Different from continuous detuning sequence from STA, here we allow 20 discretized detuning pulses for driving the system to its target quantum state. In each episode, the agent initializes the state, choosing an action after a time step, driving the state to the next time step by master equation, rewarded/punished by the environment iteratively, and evaluating the accumulated reward at the end. Parameters of the ANN updates by a given learning rate after several episodes, known as batch size. We pre-train the agent by $r(t_i)=-|\tilde{\Delta}(t_i)-\frac{i-1}{N-1}|$ for the digital pulses of LZ type , accelerating the learning process, and avoiding local optimal solutions. After the pre-training, we reward the agent by a constant value at the end of each episode if $|\rho_{22}|$ is larger than a threshold of 0.997, where a uniformly randomized systematic error within $[-0.2,0.2]$ is included in the environment within each episode as fine-tuning. We show how DRL is applied in studying only-$\sigma_z$-control by Fig.~\ref{fig:b}, finding out that either STA or DRL can lead to robust only-$\sigma_z$-control by continuous/discrete detuning. Now we look into discovering control protocols for more general cases via DRL without any knowledge of STA. In Fig.~\ref{fig:c}, we show how the effect of hybrid systematic error is minimized during an arbitrary evolving time by digital pulses. $\lambda_\Delta$ and $\lambda_\Omega$ are randomized uniformly within $[-0.1,0.1]$ in the fine-tuning process. We achieve a minimal population of $0.97$ in these area, and larger than $0.999$ in the center. We reckon that this model is well-trained, especially with the assumption that systematic errors follow Gaussian distributions.

Instead of reviewing the PPO algorithm by technical details, we here briefly introduce PPO's enhancement among others. Policy gradient methods meet convergence problems which can be solved by natural policy gradient, requiring enormous computational resources for calculating the second-order derivative matrix. PPO deals with constraints by introducing a penalty in the objective function, allowing only first-order derivative calculation for optimization. Even though it breaches constraints during training, computation is simplified with negligible damage. With clipped objective, its performance can even be better, although this argument is questioned by a recent research~\cite{ppoq}. In one word, PPO is a quick and valid DRL algorithm for our application in quantum control. 

\begin{figure}
\includegraphics[width=18cm]{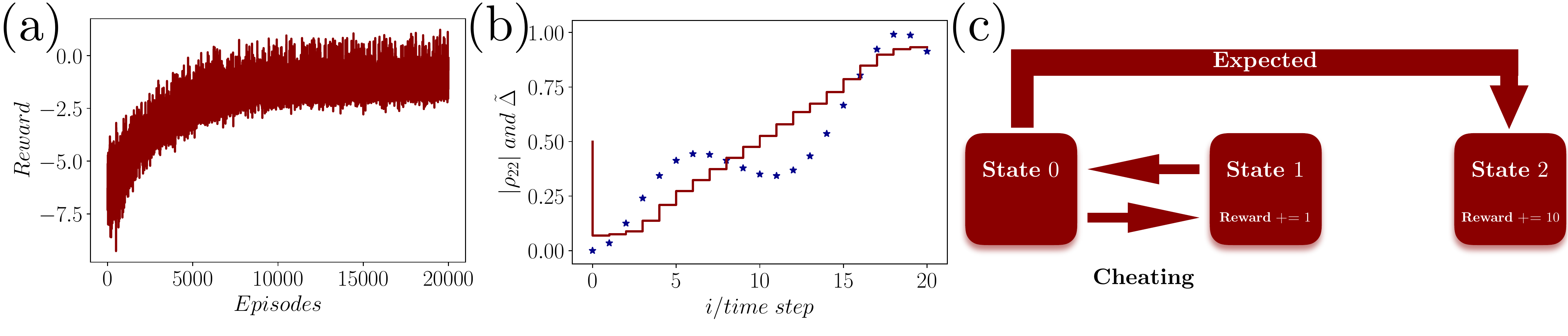}
\caption{\label{fig:sup3}(a) Reward of each episode (b) The agent for fine-tuning after pre-training. (c) Scheme of cheating by repetitive actions.}
\end{figure}

We used a DRL toolbox called TensorForce~\cite{tf} for our quick implementation. The toolbox is based on TensorFlow, which is a well-known framework for deep learning, supporting GPU acceleration. We customize a DRL environment of our two-level energy system with the combination of QuTiP~\cite{qutip}. During both pre-training and fine-tuning processes, we set a batch size of 20, and the learning rate to be 1e-4. Other hyperparameters are the default of PPO agent provided by Tensorforce. Agent learns LZ-type as our constraint faster during our pre-training with extra constant reward in the first and final timestep [see Fig.~\ref{fig:sup3}(a)]. However, one should be very careful when these tricks are used, since these extra rewards (also known as reward shaping) change the expectation of total rewards. This might change the optimal policy, resulting in cheating for more reward by repetitive action [see Fig.~\ref{fig:sup3}(c)].

Codes are compatible with both CPU and GPU version of TensorFlow 1.13.1., in Mac OS with 8-core Intel Xeon W processor and Ubuntu with Tesla P100. Codes and data are available from corresponding authors upon reasonable request.

\end{document}